\documentclass[12pt,a4paper]{article}
\usepackage{graphicx}
\usepackage{amsmath}
\usepackage[latin2]{inputenc}
\usepackage{subfigure}
\usepackage{cite}
\usepackage[normalem]{ulem}

\begin{document}

\begin{center}{\bf \Large Communities in networks - a continuous approach}\\[5mm]

{\large Ma{\l}gorzata J. Krawczyk and Krzysztof Ku{\l}akowski \\[3mm]

\em {Faculty of Physics and Applied Computer Science, AGH University of Science and Technology, al. Mickiewicza 30, 30-059 Krak\'ow, Poland\\

E-mail: kulakowski@novell.ftj.agh.edu.pl\\\today}}

\end{center}

\begin{abstract}
A system of differential equations is proposed designed as to identify communities in weighted networks. The input is a symmetric connectivity matrix $A_{ij}$. A priori information on the number of communities is not needed. To verify the dynamics, we prepared sets of separate, fully connected clusters. In this case, the matrix $A$ has a block structure
of zeros and units. A noise is introduced as positive random numbers added to zeros and subtracted from units. The task of the dynamics is to reproduce the 
initial block structure. In this test, the system outperforms the modularity algorithm, if the number of clusters is larger than four. 
\end{abstract}



\section{Introduction}

A lot of human relations in the world can be described with networks \cite{r1,r2,r3,r4,r5,r6,r7,r8,bocc,dorgo}. 
We ourselves are elements of 
social networks connecting people in the same family, faculty or club. 
We can also be involved in networks through the Internet - we send and receive e-mails, 
we create WWW pages, and we visit a lot of them every day.
Being employed, we are connected with people with the same field of interests. These connections are 
sometimes surprisingly far reaching especially nowadays when interdisciplinary tasks 
are very common. Other social networks are formed with religious or political ties. 
Another family is formed by economical networks; financial positions of many firms are connected to each other. 
 Knowledge of the connections 
between firms enable an efficient diversification of the portfolio or forecasting of risk. This 
kind of networks awaken people's interest because of potential possibility of enrichment, 
thanks to anticipation of the behaviour of the financial market.
 Studies of these and many 
other complex networks require not only knowledge of a given problem, e.g. sociology or 
economy, but also skills of application of the methods which enable correct analysis. 
Proper reconstruction of a network of interactions between interesting elements allows 
to predict selected aspects of the system.

In particular, the problem of identification of communities in networks is relevant for numerous areas of knowledge. Economists are interested in 
clusterization of time series of financial data \cite{mant,kim}. In biology the correlations between genes expression allow to infer
on similar functions \cite{chu,simek}. Sociologists investigate the structure of social networks, looking for groups and their leaders
\cite{fre,bon}. However, the problem is computationally difficult; time needed by exhaustive algorithms grows exponentially with the 
network size \cite{new1}. Numerous approximate solutions have been applied; for a recent review see Ref. \cite{bocc}.

In most of these works, the algorithms applied rely on discrete variables. Continuous dynamics happens to be used in networks within 
the mean-field theory, where the time evolution of a few collective variables is considered. This approach is presented e. g. in 
Ref. \cite{proulx}. Also, differential equations have been applied to describe the time evolution of nodes of random networks \cite{gomez}.
However, we are not aware on any application of the continuous dynamics to the topology itself, except the physically motivated approach 
of Wu and Huberman \cite{wu} and our own works on the Heider balance
\cite{fir,myr}. The algorithm presented in Ref. \cite{wu} relies actually on an approximated calculation of a continuous distribution of electric potential in a network. The Heider balance is a specific version of the problem because we have only two communities. 
However, as we demonstrate below, this drawback is easy to be removed by some reformulation of the model equations. In our opinion,
the continuous description has some advantages which deserve the modeller's attention. First, in many applications the continuous variables 
seem to be more natural than the discrete ones. Second, with discrete variables often we have to rely on the stochastic dynamics, what 
in principle yields averaging over trajectories even if an initial state is given. Alternatively we have to present results which cannot 
be reproduced exactly by other authors, as they use different series of pseudorandom numbers. The advantage of discrete models is that
the time of calculations is much shorter. 

In this paper we propose a set of differential equations to describe the time evolution of links in a weighted network. Initially, the 
weights of all links are given by real numbers between zero and one. The equation are designed as to drive the weights to 0 or 1;
then a link is supposed to disappear or persist. The obtained network reveals the cluster structure, which is supposed to be the 
closest to the initial one. This kind of evolution is analogous to a task to find the closest energy minimum in a complex landscape.

The question remains, how to verify a given formalism. Often we refer to social experiments, as the Zachary karate club \cite{new3,myr}. 
However, if this kind of accordance appears, it should be treated merely as a nice demonstration that the scheme of calculations is 
reasonable rather than as an ultimate proof. In social experiments, there is so many uncontrolled factors that an exact prediction of the
output is rarely possible; any theory is by necessity reductionistic, and only a limited part of reality can be taken into account.
(Being aware that usually only qualitative accordance is possible, social scientists tend to treat a perfect accordance of theory with 
experiment as somewhat suspicious). Another possibility is to compare many different approaches \cite{fre}. However, with this method we cannot find 
one best algorithm, but rather to eliminate the worst one. Having this in mind, we designed a numerical test to compare our method
with two other algorithms available in the literature, the modularity algorithm \cite{new2} and the shortest path
algorithm \cite{new3}. The test is designed as that we know the proper results; in this way the above difficulties are omitted.

In the next section we describe the model equations and the test. In Section 3 we compare the results obtained with the three methods.
Last section is devoted to conclusions.

\section{Model equations and the test method}

The input to the calculation is an initial state of the symmetric real matrix $A$. Each matrix element $A_{ij}$ refers to the state of the 
link between nodes $i$ and $j$. The time evolution of the matrix elements is determined by

\begin{equation}
\frac{dA_{ij}}{dt}=G(A_{ij})\sum_k(A_{ik}A_{kj}-\beta)
\end{equation}
where $G(x)=\Theta(x)\Theta(1-x)$. This product of the step functions ensures that the matrix elements remain in the prescribed range (0,1). 
The parameter $\beta$ is a threshold (see below). The idea is that for each link $ij$ all remaining nodes $k$ are to be considered one by one. The question 
is, if the link $ij$ joins nodes from the same cluster or from different clusters? To decide this, we consider the products $A_{ik}A_{kj}$; if both elements $A_{ik}$ and $A_{kj}$ are large, they are supposed to join the nodes
in the same cluster; then also $i$ and $j$ are in the same cluster. In this case, the product $A_{ik}A_{kj}$ is larger than the threshold 
value $\beta$, and $A_{ij}$ increases. This tendency to increase or decrease is averaged over all nodes $k$. As a result, the matrix elements
vary with different velocities; evident parts of the cluster structure are determined at first, and they determine the further evolution.

In our calculations, the parameter $\beta$ is fixed as 0.25. This choice is motivated as follows. Our aim is to discriminate two cases:
{\it i)} when nodes $i$ and $k$ belong to the same cluster and nodes $k$ and $j$ belong to two different clusters, {\it ii)} when all three
nodes belong to the same cluster. A priori, $A_{ik}$ and $A_{kj}$ can be treated as independent random variables; if their distribution
is uniform in the range (0,1), the average of their product is $(1/2)^2$.

To check how the method works, we use a set of $R$ identical, fully connected clusters ("cliques" \cite{bocc}), each of $K$ nodes; the whole number of nodes is $N=RK$. The $N\times N$ connectivity matrix of this system
has a block structure, with squares $K\times K$ of units along the main diagonal and zeros elsewhere. This initial structure is hidden
when all matrix elements are disturbed; for each matrix element $A_{ij}$ a random number $\epsilon_{ij}$ is generated with a uniform distribution
from the range ($0,e$). The transformation is $A_{ij}\to B_{ij}(A_{ij},\epsilon_{ij})$, where

\begin{equation}
B(x,\epsilon)=(1-x)\epsilon+x(x-\epsilon)
\end{equation}
In this way, $\epsilon$ is added to zeros and subtracted from units. This noised connectivity matrix $B$ serves as an initial 
condition to Eq. (1). The time evolution should reproduce structure of the initial matrix $A_{ij}$ at the maximum of the modularity $Q$, defined as \cite{new2}

\begin{equation}
Q=\frac{1}{w}\sum_{ij}(B_{ij}-\frac{k_ik_j}{w})\delta(c_i,c_j)
\end{equation}
where $w=\sum_{ij}B_{ij}$, $k_i=\sum_jB_{ij}$ and the factor $\delta$ indicates that only nodes in the same cluster are taken into 
account. This quantity allows to distinguish the network with given structure from the equivalent random network.

The same $B_{ij}$ matrix is the input for the modularity algorithm \cite{new2} and the algorithm of shortest paths \cite{new3}. In the case of the weighted network the shortest paths method requires a redefinition of the distance matrix. Our tests indicate that the 
best results are obtained if the non-diagonal elements of the matrix are equal to reverse square of the correlation 
coefficient. In order to find all shortest paths connecting pairs of nodes in the network the modified Bellman-Ford 
algorithm was used \cite{algo}. Modification of the algorithm was necessary because the original version does not 
take into account all possible shortest paths. At each iteration just this edge is removed from the 
network and the modified Bellman-Ford algorithm is applied to the obtained network. The simulation is terminated 
when all edges are removed from the system.

\section{Results}

In Figs. 1-7 we show the results of the calculations of the probability $p$, that the initial connectivity matrix $A_{ij}$ will be reproduced.
In all figures, the curves are marked as follows: triangles for Eq. 1, crosses for the modularity and squares for the shortest paths.
The results are shown for various $N$ and $R$.
The probability {p} depends on the maximal value $e$ of the random numbers, used to introduce a noise. All results are averaged over 100 matrices. It is obvious, that for small $e$ all 
methods work well; there, $p$ is equal to unity for all three methods. However as $e$ increases, errors appear and the probability $p$ 
starts to decrease. 

\begin{figure}
\vspace{0.3cm} 
{\par\centering \resizebox*{10cm}{8cm}{\rotatebox{0}{\includegraphics{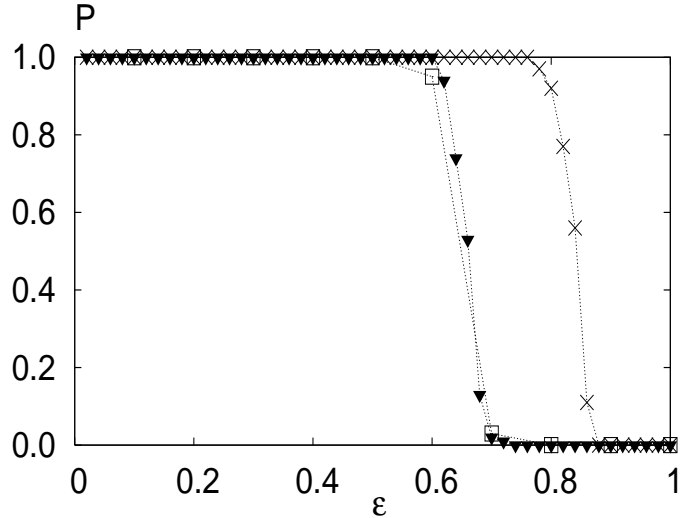}}} \par} 
\vspace{0.3cm} 
\caption{Results for $R$=2 clusters, each of $K$=34 nodes. The modularity algorithm works best.}  
\end{figure}

\begin{figure}
\vspace{0.3cm} 
{\par\centering \resizebox*{10cm}{8cm}{\rotatebox{0}{\includegraphics{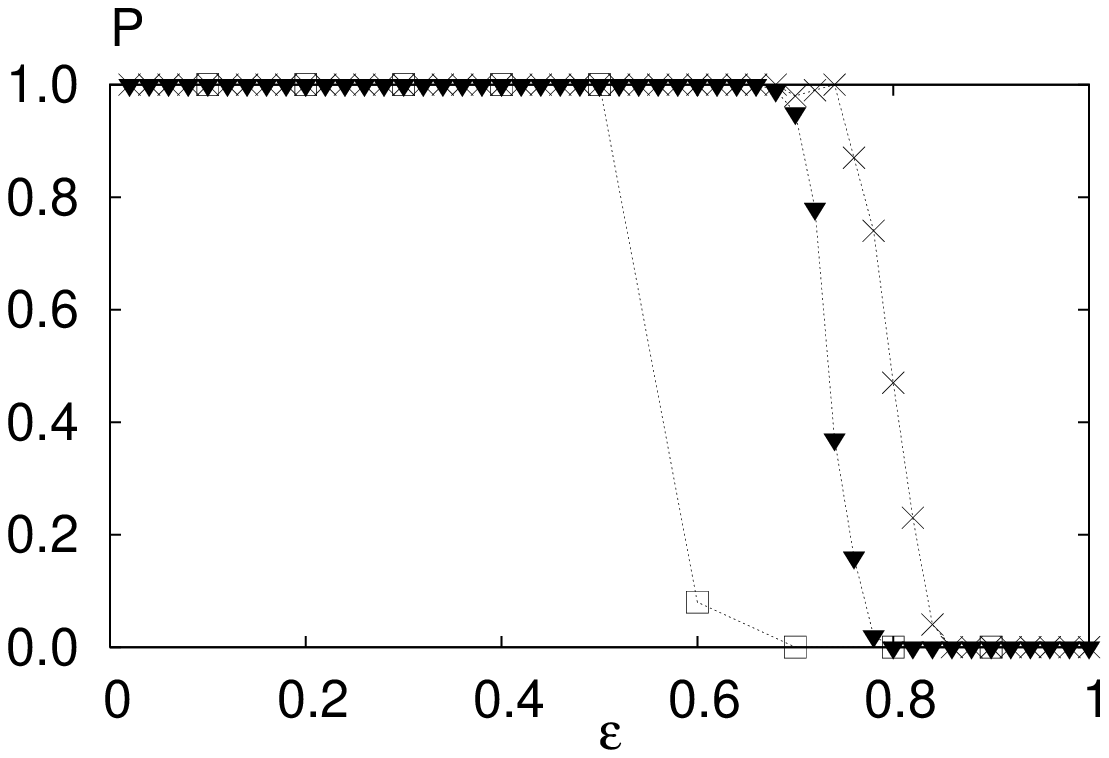}}} \par} 
\vspace{0.3cm} 
\caption{Results for $R$=3 clusters, each of $K$=24 nodes. The modularity algorithm works best.}  
\end{figure}

\begin{figure}
\vspace{0.3cm} 
{\par\centering \resizebox*{10cm}{8cm}{\rotatebox{0}{\includegraphics{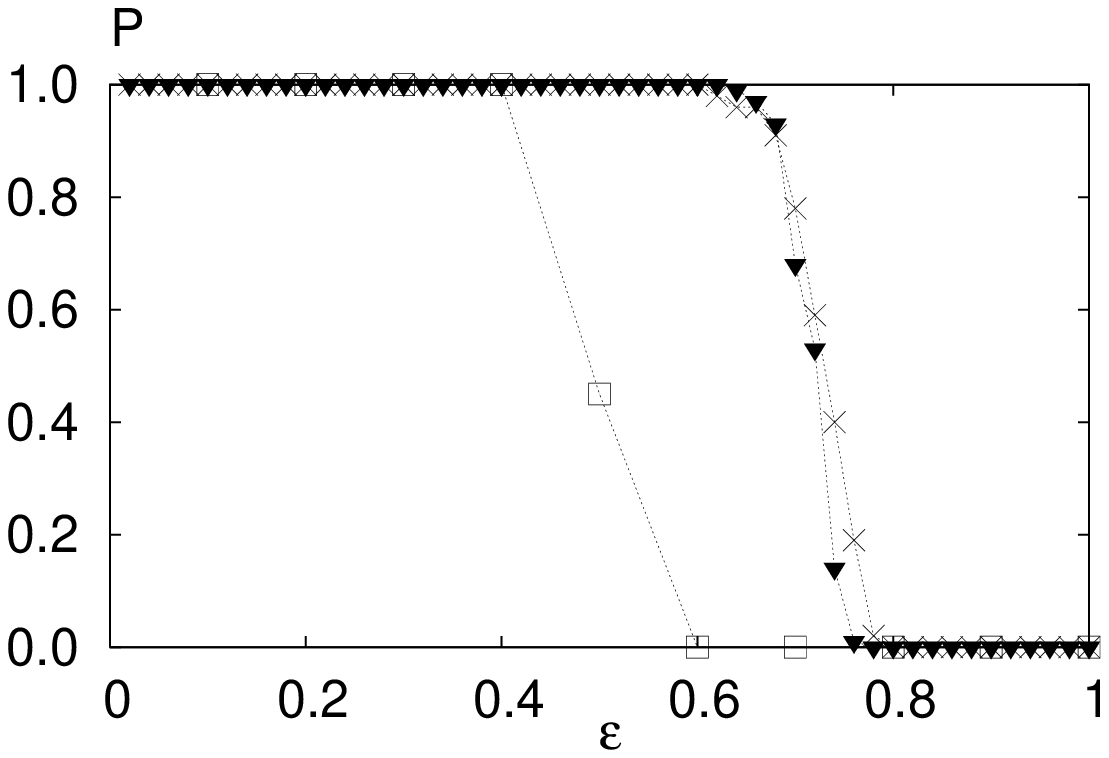}}} \par} 
\vspace{0.3cm} 
\caption{Results for $R$=4 clusters, each of $K$=15 nodes. The results of Eq. 1 are as good as those of the modularity algorithm.}  
\end{figure}

\begin{figure}
\vspace{0.3cm} 
{\par\centering \resizebox*{10cm}{8cm}{\rotatebox{0}{\includegraphics{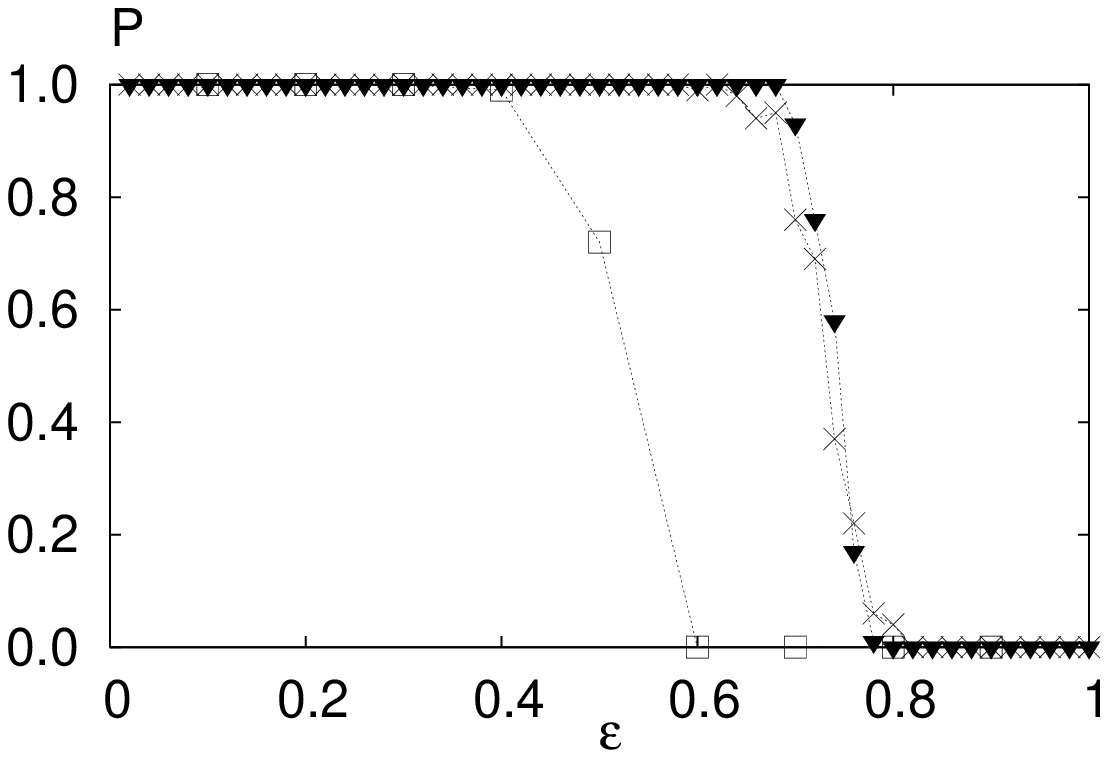}}} \par} 
\vspace{0.3cm} 
\caption{Results for $R$=4 clusters, each of $K$=20 nodes. The results of Eq. 1 are as good as those of the modularity algorithm.}  
\end{figure}

\begin{figure}
\vspace{0.3cm} 
{\par\centering \resizebox*{10cm}{8cm}{\rotatebox{0}{\includegraphics{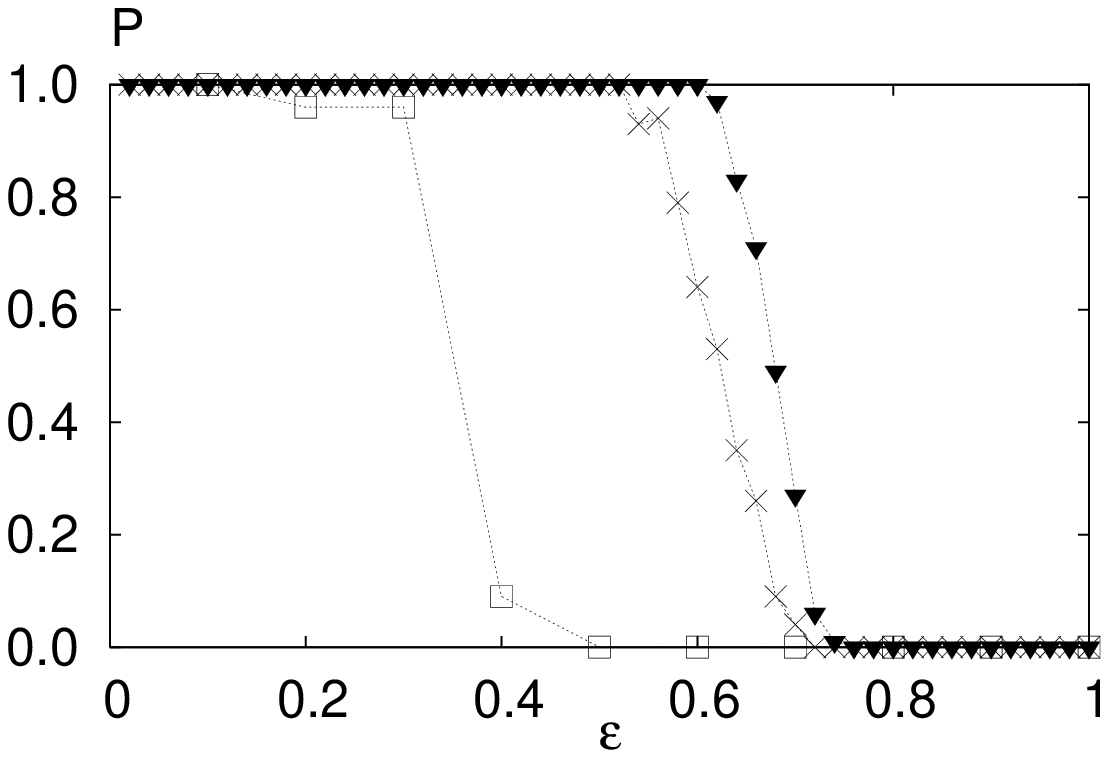}}} \par} 
\vspace{0.3cm} 
\caption{Results for $R$=6 clusters, each of $K$=10 nodes. Eq. 1 works best.}  
\end{figure}

\begin{figure}
\vspace{0.3cm} 
{\par\centering \resizebox*{10cm}{8cm}{\rotatebox{0}{\includegraphics{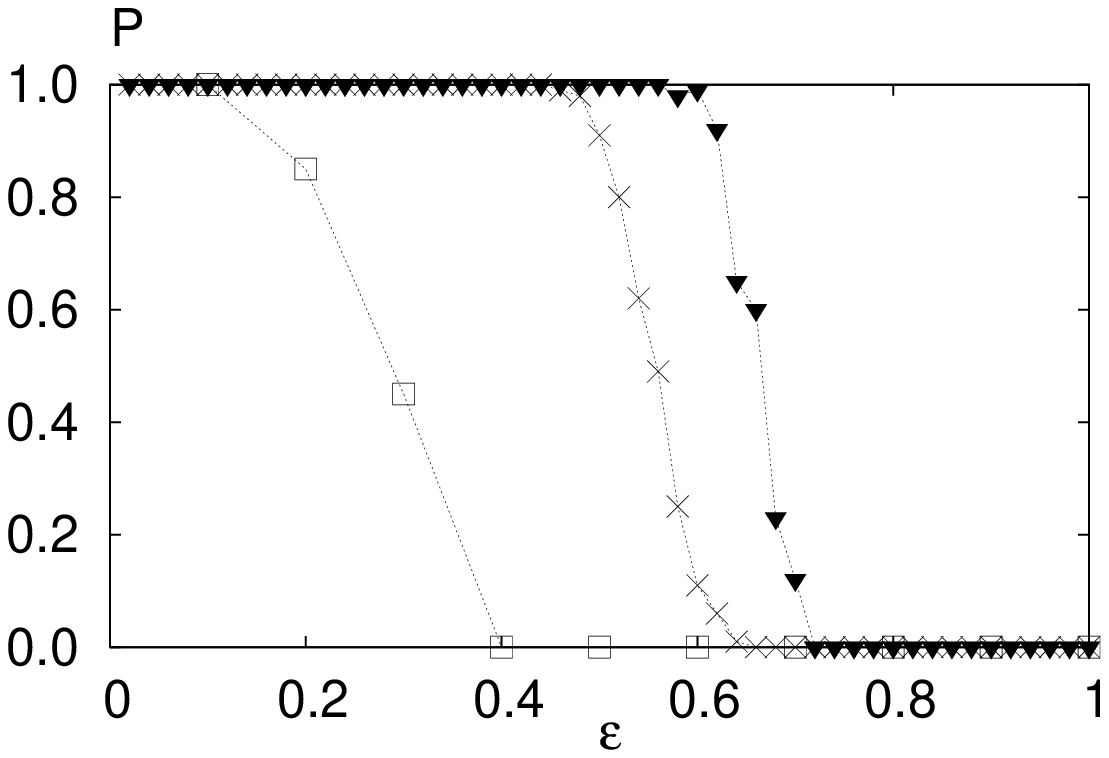}}} \par} 
\vspace{0.3cm} 
\caption{Results for $R$=8 clusters, each of $K$=9 nodes. Eq. 1 works best.}  
\end{figure}

\begin{figure}
\vspace{0.3cm} 
{\par\centering \resizebox*{10cm}{8cm}{\rotatebox{0}{\includegraphics{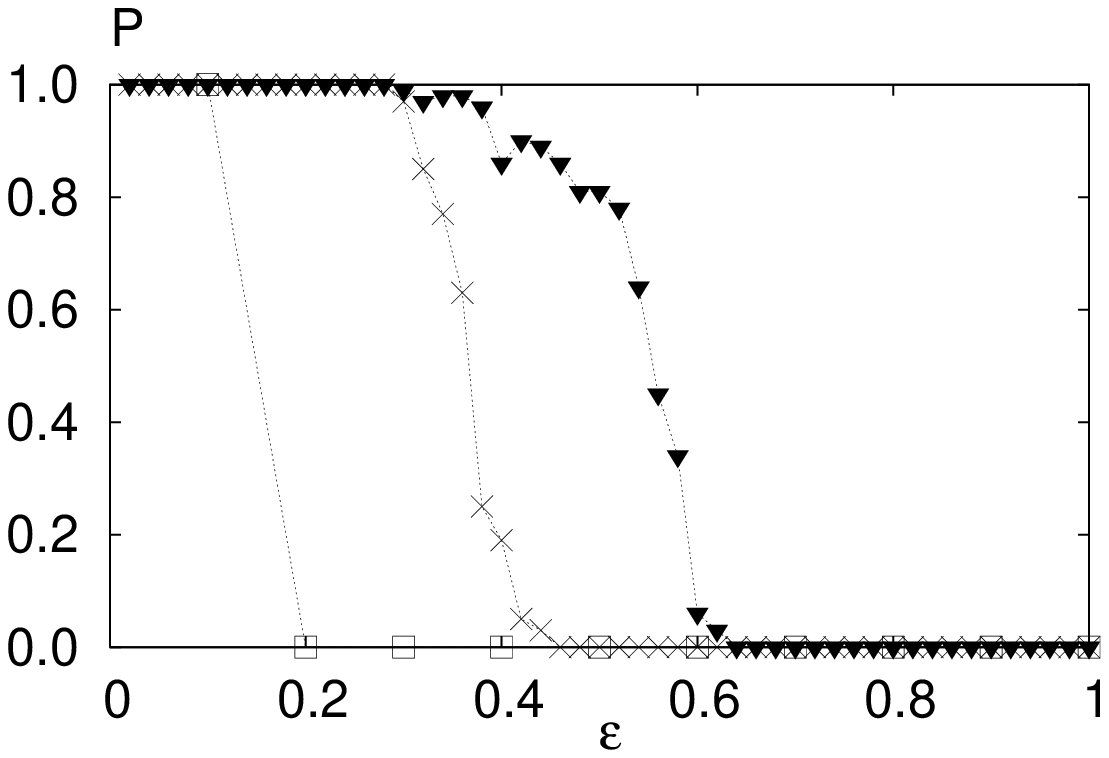}}} \par} 
\vspace{0.3cm} 
\caption{Results for $R$=16 clusters, each of $K$=5 nodes. Eq. 1 works best.}  
\end{figure}

As we see, the shortest path method fails for the smallest noise. The modularity method is the best for $R$=2 and 3. For $R$=4, the modularity
method and Eq. 1 give approximately the same results. For $R$=6 and higher, the probability of success is the largest for the evolution given by Eq. 1.

\section{Discussion}

Our results are preliminary. Further calculation should explore the parameter space ($R,K$) more thoroughly. The method should also be 
checked for clusters of different sizes, which are not fully connected. As the number of internal links is reduced, the clusters are "less dense". When the number of strong links between such clusters is too large, the identification of communities becomes impossible - this is the natural boundary of any method and of the definition of communities as well. It will be interesting to repeat 
our comparison in this difficult area. The drawback is the time of computation; the number of differential equations to be solved numerically is $N(N-1)/2$. This limitation can be somewhat returned by time of writing the code, as the method is conceptually very simple. Our results indicate, 
that it can be useful when applied to networks of moderate size, but with larger number of clusters. Also, it can be immediately generalized to unweighted networks; in this case the links are described by integers at the beginning and at the end of the calculation.

As it can be seen in the presented figures, the method works well in the case of large number of clusters, but its efficiency is worse for two or three clusters. In the latter case the number of triads $i,j,k$ where nodes $i,j$ belong to the same cluster but $k$ does not is relatively large. The product $A_{ik}A_{kj}$ is lower than the threshold $b$ and these triads produce a decrease of $A_{ij}$; this 
lowers the performance of the whole method when the number of clusters is small.

There is some relation between the problem of communities and the infinite-range spin glass problem. However, this relation cannot be reduced to an identity. For two communities, a spin can be defined as
a variable which marks to which community a node does belong.
However, for a larger number of communities this analogy does not hold. The set of potentially stable solutions of the continuous dynamics contains in particular the case when every node belongs to a different cluster. Then, if we try to apply the spin analogy, the spin dimension should be at least equal to the number of nodes. It seems that it is more proper to work with the variables denoting the states of the links only.

\end{document}